\documentclass[aps, prl, reprint, superscriptaddress,preprintnumbers,amssymb]{revtex4-2}
\usepackage{graphicx}% Include figure files
\usepackage{dcolumn}% Align table columns on decimal point
\usepackage{bm}% bold math

\begin{document}

\preprint{ver4p1}

\title{Fermi-surface-sheet dependent electron-phonon coupling in a borocarbide superconductor YNi$_2$B$_2$C}


%\author{}
%\thanks{}
%\affiliation{}

\author{Taichi Terashima}
\email{TERASHIMA.Taichi@nims.go.jp}
\affiliation{Research Center for Materials Nanoarchitectonics (MANA), National Institute for Materials Science, Tsukuba 305-0003, Japan}
\author{Hiroyuki Takeya}
\affiliation{Research Center for Energy and Environmental Materials (GREEN), National Institute for Materials Science, Tsukuba 305-0047, Japan}
\author{Hisatomo Harima}
\affiliation{Department of Physics, Kobe University, Kobe 657-8501, Japan}

%\no affiliation

\date{\today}

\begin{abstract}
We performed de Haas-van Alphen (dHvA) oscillation measurements and band-structure calculations for YNi$_2$B$_2$C. 
Our improved band structure successfully explained the origins of the large dHvA frequencies $\beta$ and $\zeta$, which were inexplicable in previous works. 
By comparing experimental effective masses with band masses, we determined the electron-phonon coupling for each orbit. 
The results showed a clear Fermi-surface-sheet dependence of the electron-phonon coupling strength, especially highlighting that the coupling for the band-28 sheet is very weak, almost absent for the orbit with $B \parallel c$. 
This finding is consistent with previous observations of dHvA oscillations from this orbit in the mixed state down to very low fields. 
Amidst growing interest in high-temperature superconductivity driven by electron-phonon coupling in hydrides under high pressure, this study provides foundational data pivotal to precisely understanding electron-phonon coupling.
\end{abstract}

%\keywords{}

\maketitle

%\section{}
%\subsection{}
%\subsubsection{}

% If in two-column mode, this environment will change to single-column
% format so that long equations can be displayed. Use
%\begin{widetext}
%\end{widetext}

% figures should be put into the text as floats.
% Use the graphics or graphicx packages (distributed with LaTeX2e)
% and the \includegraphics macro defined in those packages.
% See the LaTeX Graphics Companion by Michel Goosens, Sebastian Rahtz,
% entire page. There is no need to do explicit centering.

de Haas--van Alphen (dHvA) oscillation measurements combined with band-structure calculations are a powerful tool to investigate bulk electronic structures in metals.
One can determine not only the shape of the Fermi surface but also the effective mass.
The effective mass $m^*$ determined by dHvA measurements is enhanced over the band mass $m_{band}$ by electron-phonon and electron-electron interactions: $m^*=(1+\lambda)m_{band}$.
Thus one can characterize the many-body interactions in metals by determining the enhancement parameter $\lambda$ for different cyclotron orbits on different Fermi-surface sheets.
Pioneering work was conducted nearly 40 years ago on the electron-phonon coupling superconductor Nb, revealing that the strength of electron-phonon coupling varies across different sheets of the Fermi surface with up to $\pm40\%$ variation \cite{Crabtree87PRB}. 
More recently, it has been reported that the magnitude of electron-phonon coupling in MgB$_2$ differs significantly between the $\sigma$ and $\pi$ bands \cite{Yelland02PRL, Carrington03PRL}.

DHvA oscillations can also be observed in the mixed state of superconductors: e.g., NbSe$_2$ \cite{Graebner76PRL, Onuki92JPSJ}, A15 superconductors \cite{Mueller92PRL, Corcoran94PRL, Harrison94PRB}, borocarbide superconductors \cite{Terashima95SSC, Heinecke95ZPB, Bergk07PhysicaC, Isshiki08PRB}, organic superconductors \cite{vanderWel95SynthMet, Wosnitza00PRB}, heavy-fermion superconductors \cite{Hedo95JPSJ, Ohkuni97JPSJ, Haga99JPSJ, Settai01JPCM, Terashima07PRB}, MgB$_2$ \cite{Yelland02PRL, Fletcher04PRB, Isshiki05PhysicaC}, KOs$_2$O$_6$ \cite{Terashima12PRB}, and high-$T_c$ cuprates \cite{DoironLeyraud07Nature, Sebastian08Nature}.
dHvA oscillations in the mixed state suffer extra damping compared to the normal state. 
This damping is due to the opening of the superconducting gap, and therefore, by analyzing this damping, it is expected that one can estimate the magnitude of the superconducting gap for each orbit, although, at present, many different theoretical models have been proposed \cite{Maki91PRB, Stephen92PRB, Maniv92PRB, Miyake93PhysicaB, Wasserman94PhysicaB, Dukan95PRL, Norman95PRB, Miller95JPCM, Gvozdikov98PRB, Gorkov98JETPLett, Mineev00PhilMagB, Yasui02PRB, Zhuravlev12PRB, Gorkov12PRB}, and it is unclear which model is the most quantitatively reliable (see, for example, discussion in \cite{Janssen98PRB, Maniv01RMP, Yasui02PRB}).

In this paper, we focus on YNi$_2$B$_2$C, a nonmagnetic member of the rare-earth nickel-borocarbide superconductors \cite{Nagarajan94PRL, Cava94Nature}, with a relatively high superconducting transition temperature of $T_c$ = 15.6 K.
Although those borocarbides contain a typical magnetic element Ni, phonon-mediated superconductivity was theoretically suggested \cite{Pickett94PRL} and finally confirmed experimentally by the observation of clear isotope effects \cite{Cheon99PhysicaC}.
Thus the mass enhancement in YNi$_2$B$_2$C is basically ascribed to the electron-phonon coupling.
%The electron-phonon interaction in YNi$_2$B$_2$C has been extensively studied both theoretically and experimentally \cite{Reichardt05JSupercond,Weber08PRL, Weber12PRL,Weber14PRB,Tutuncu15PRB,Kawamura17PRB, Kurzhals22NatCommun, Christiansson24PRB}.
Intriguingly, despite the phonon-mediated superconductivity, strong anisotropy in the superconducting gap has been suggested from upper critical field, specific heat, thermal conductivity, and angle-resolved photoelectron spectroscopy measurements \cite{Shulga98PRL, Nohara99JPSJ, Boaknin01PRL, Baba10PRB}.

Several research groups have conducted dHvA measurements on YNi$_2$B$_2$C and a sister compound LuNi$_2$B$_2$C to determine the Fermi surface \cite{Terashima95SSC, Heinecke95ZPB, Tokunaga95JPSJ, Nguyen96JLTP, Terashima97PRB, Bergk08PRL, Isshiki08PRB}.
Although comparison to band-structure calculations was partially successful, there was a notable discrepancy:
the calculations were unable to explain the origin of the second largest frequency $\beta$ with $F \sim$7 kT for $B \parallel c$ (Fig. 1) \cite{Winzer01Book, Yamauchi04PhysicaC, Bergk08PRL, Isshiki08PRB}.
In addition, a large frequency $\zeta$ remained unexplained as well (Fig. 1).
Because the $\beta$ orbit occupies more than 20\% of the cross-sectional area of the Brillouin zone, these failures are serious.
Intriguingly, the $\alpha$ oscillation arising from a small closed Fermi sheet in YNi$_2$B$_2$C could be observed deep in the mixed state, in some cases down to a field as low as $\sim$20\% of the upper critical field (0.2$B_{c2}$) \cite{Terashima95SSC, Heinecke95ZPB, Goll96PRB, Terashima97PRB, Bintley03PhysicaC, Isshiki08PRB, Nossler17PRB}.
It was argued that the gap opening on this sheet is much smaller than gaps on other sheets \cite{Terashima97PRB, Nossler17PRB}.

In this study, we perform dHvA measurements on YNi$_2$B$_2$C for various field directions.
We show that our improved band-structure calculation successfully explains the origins of the previously inexplicable $\beta$ and $\zeta$ frequencies.
We then compare the experimental effective masses with band masses to determine the mass enhancement parameter $\lambda$ for each orbit.
The results clearly show the Fermi-surface-sheet dependence of the mass enhancement.
Especially, the mass enhancement for the $\alpha$ orbit is very small, nearly absent for $B \parallel c$.
This strongly suggests that the $\alpha$ sheet is passive for superconductivity and hence that the gap on this sheet is small, in line with the observation of dHvA oscillations from the $\alpha$ sheet at very low fields in the mixed state.
Although the mass enhancement in LuNi$_2$B$_2$C was previously determined in a similar manner \cite{Bergk08PRL}, the band-structure used in the study was unable to explain the $\beta$ frequency and also the $\phi$ frequency (Fig. 1), which restricts the reliability of the obtained results.
The recent discovery of high-temperature superconductivity in hydrides under high pressure has sparked renewed interest in the electron-phonon interaction \cite{Drozdov15Nature, Drozdov19Nature}.
The precise determination of the Fermi-sheet dependent electron-phonon coupling demonstrated in this study makes an important contribution to such a research trend.

YNi$_2$B$_2$C single crystals were grown by a floating zone method \cite{Takeya96JAC, SM}.
Field-modulation dHvA measurements were performed in a dilution refrigerator installed in a 16-T or a 20-T superconducting magnet \cite{*[][ for details of the crystal growth{,} characterization{,} dHvA measurements{,} band-structure calculations{,} and estimation of $\lambda${,} which includes Refs{.} 79--82{.}] SM}.
The magnetic field was rotated in the (1$\bar{1}$0), (010), and (001) planes.
The field angles $\theta_{(1\bar{1}0) \,\mathrm{or}\, (010)}$ and $\varphi$ are measured from the [001] and [100] axes, respectively.
Four samples were measured and gave consistent results.

\begin{figure}
\includegraphics[width=8.6cm]{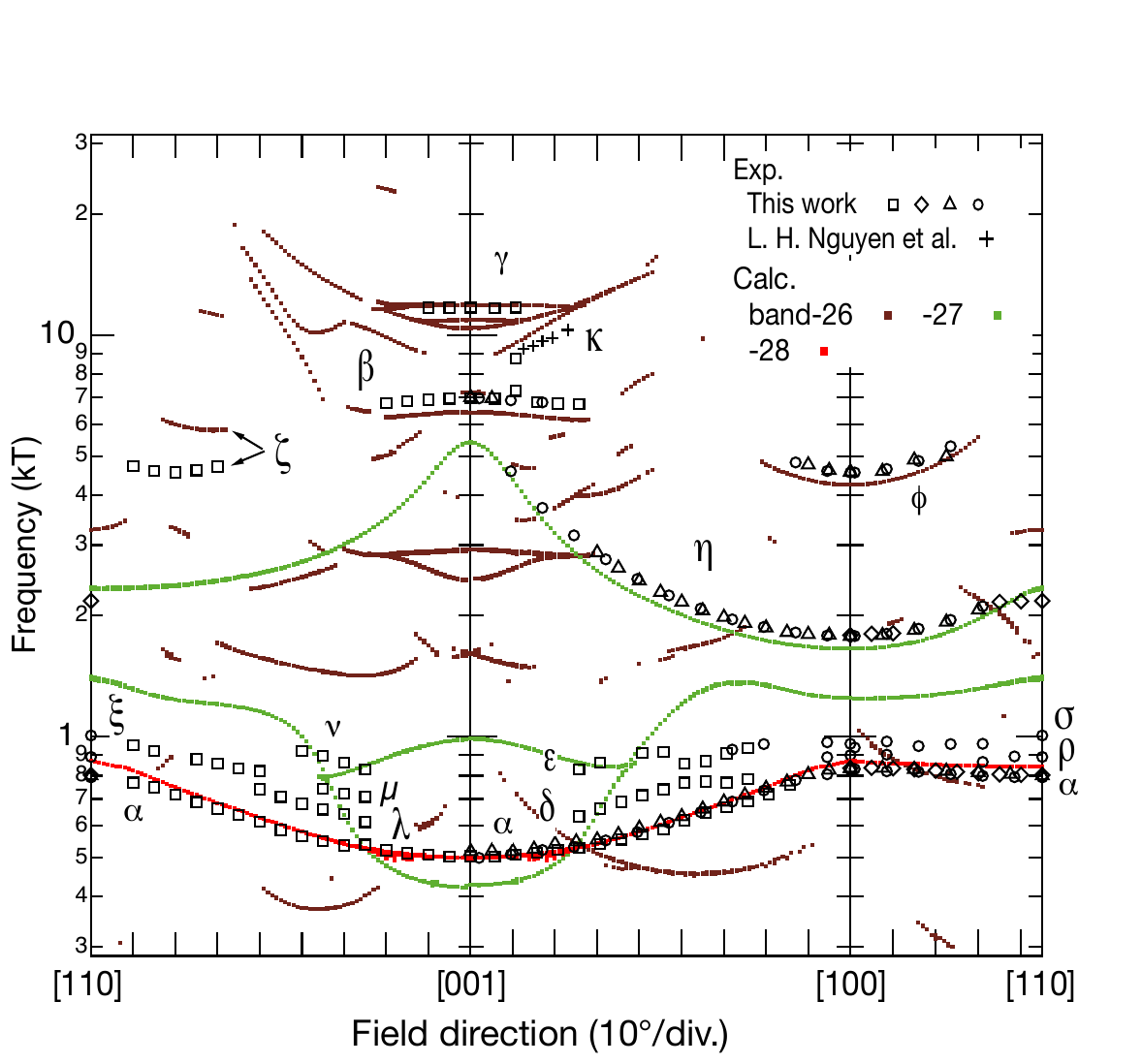}
\caption{\label{Fig1}Experimental vs theoretical dHvA frequencies in YNi$_2$B$_2$C.
Squares, diamonds, triangles, and circles indicate dHvA frequencies experimentally observed in the present four samples of YNi$_2$B$_2$C.
Crosses are from \cite{Nguyen96JLTP}.
Dots indicate frequencies calculated from the present band-structure calculation.
}
\end{figure}

\begin{figure}
\includegraphics[width=7cm]{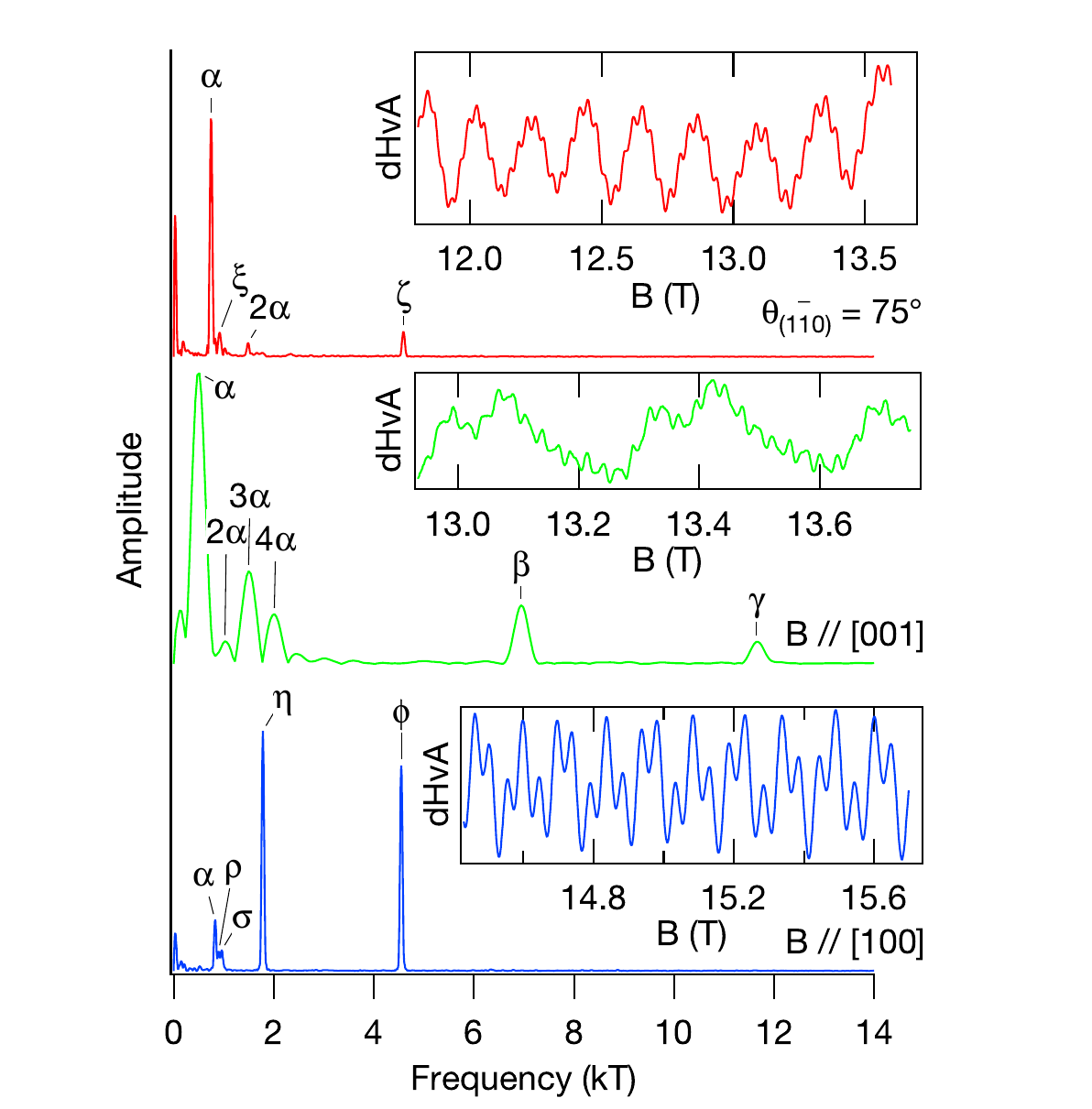}
\caption{\label{Fig2}de Haas--van Alphen oscillation in YNi$_2$B$_2$C.
dHvA oscillations detected by field-modulation technique (inset) and corresponding Fourier transforms are shown for indicated magnetic field directions.
The measurement temperatures were 30--50 mK.}
\end{figure}

Figure 2 shows examples of the observed dHvA oscillations and corresponding Fourier transforms.
Figure 1 plots the observed dHvA frequencies as a function of the field direction.
According to the Onsager relation, the frequency of the dHvA oscillations is proportional to the $k$-space area of the observed cyclotron orbit \cite{Shoenberg84}. 
In the present case, for $B \parallel c$, the frequency $F$ = 300 T corresponds to an orbit whose area is about 1\% of the cross-sectional area of the Brillouin zone.
Many frequency branches were observed as labelled by Greek letters.
The frequencies observed in the four samples agree well when the measurement directions overlap.
The observed field-angle dependence of the frequencies is in line with previously reported ones \cite{Terashima95SSC, Nguyen96JLTP}.
It is also very similar to that reported for LuNi$_2$B$_2$C \cite{Bergk08PRL, Isshiki08PRB}.

In order to interpret the observed frequency branches, we calculated the electronic band structure within the local density approximation (LDA), using a full-potential linearized augmented plane-wave (FLAPW) method \cite{SM}.
Following \cite{Yamauchi04PhysicaC}, we shifted the Y $d$ and Ni $d$ levels upward from the LDA levels by 0.11 and 0.05 Ry, respectively.
Empirically, such a shift is known to improve the agreement between the calculated Fermi surface and experimental observations \cite{Harima88SSC, Ebihara00JPSJ, Harima04JMMM, Matsuda05JPSJ, Nakamura14JPSJ}.
This indicates that the relative energy levels of electrons obtained by LDA should sometimes be corrected.
We used the experimental lattice parameters: $a$ = 3.526 \AA, $c$ = 10.543 \AA, and $z$(B) = 0.358 \cite{Godart95PRB}.
The present calculation is basically the same as that in \cite{Yamauchi04PhysicaC}, but a much finer $k$-mesh ($24 \times 24 \times 8$; 767 $k$-points in the irreducible Brillouin zone) was used to capture fine details of the Fermi surface.

\begin{figure}
\includegraphics[width=8.5cm]{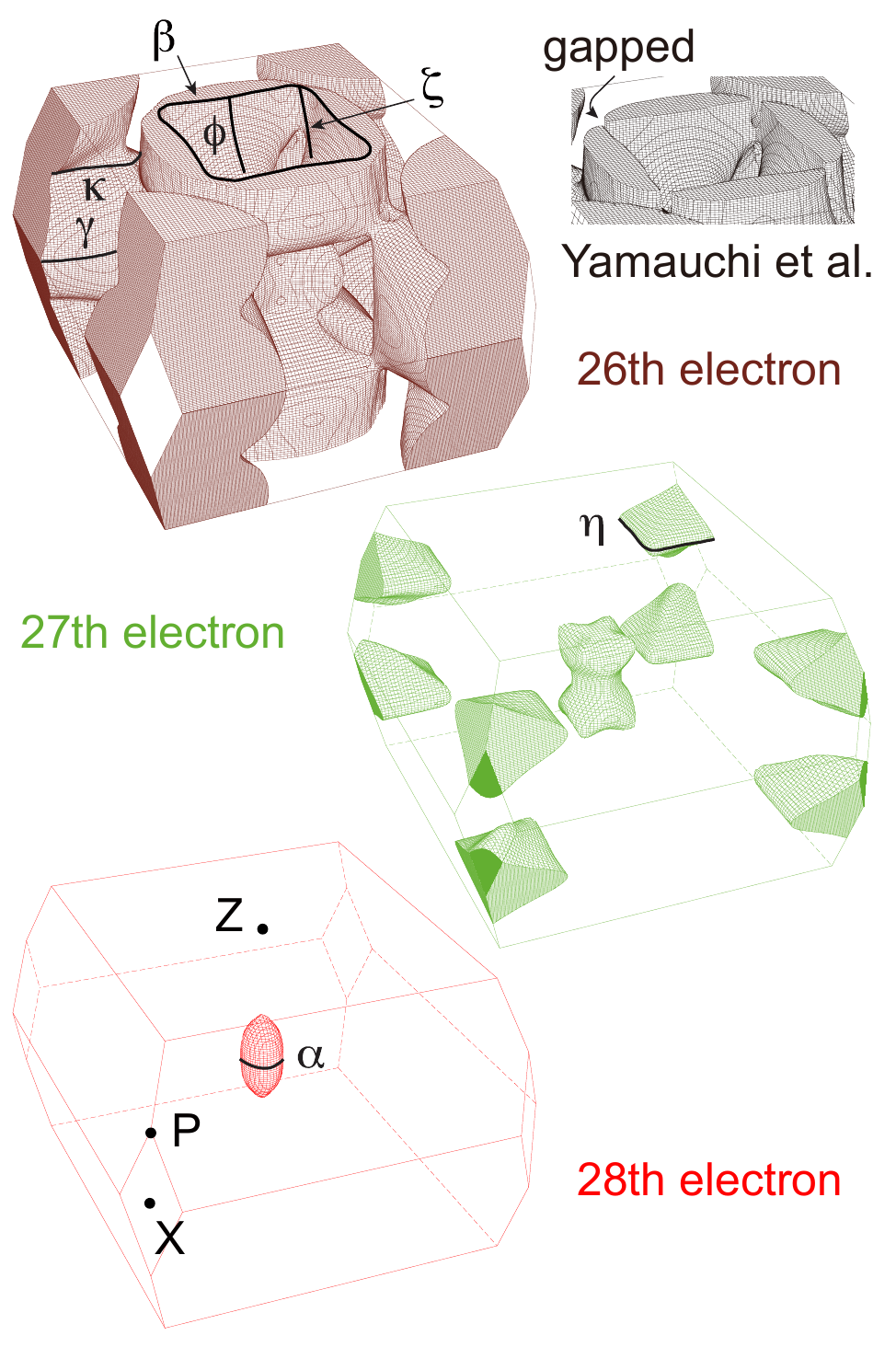}
\caption{\label{Fig2}Calculated Fermi surface in YNi$_2$B$_2$C.
Extremal orbits are indicated.
Top right inset shows part of the band-26 electron surface calculated in \cite{Yamauchi04PhysicaC}, for comparison.
}
\end{figure}

Three bands cross the Fermi level [Fig. S3(a)], and the calculated density of states is 58.3 states/Ry.
This corresponds to the electronic specific heat coefficient of $\gamma_{band}$ = 10.1 mJ/mol~K$^2$, and, compared to an experimental value of $\gamma_{exp}$ = 18.2 mJ/mol~K$^2$ \cite{Michor95PRB}, gives a Fermi-surface averaged electron-phonon coupling of $\langle \lambda \rangle$ = $\gamma_{exp}/\gamma_{band} - 1$ = 0.8.
Figure 3 shows the calculated Fermi surface.
The 26th band produces the multiply-connected electron sheet, the 27th band a closed electron sheet at $\Gamma$ and closed ones at P, and the 28th a closed electron sheet at $\Gamma$.

We computed dHvA frequencies based on the calculated band structure (Fig. 1).
In the calculation, we shifted the 28th band downward by 5 mRy (68 meV) to achieve a better agreement with the $\alpha$ frequency.
Such a small adjustment of the band energy ($\sim100$ meV) is a common practice when comparing theoretical and experimental dHvA frequencies \cite{Carrington05PRB, Coldea08PRL, Analytis09PRL, Terashima11PRL, Baglo22PRL}.
This necessity may arise from numerical errors in the calculations or from electronic correlations that are not sufficiently taken into account within LDA.
As seen from Fig. 1, our Fermi surface can explain most of the experimental frequency branches, i.e., $\alpha$, $\beta$, $\gamma$, $\zeta$, $\eta$, $\kappa$, and $\phi$, whose corresponding orbits are depicted in Fig. 3 [for $\zeta$ and $\phi$, see also Fig. S3(c)].
For the $\zeta$ branch, there is a slight difference between the experimental and calculated frequencies.
However, this frequency difference corresponds to only a small energy shift of about 9.6 mRy (130 meV) within an effective-mass approximation.
(We did not shift the 26th band because it would affect multiple frequency branches.)
The remaining branches may be assigned to the band-27 $\Gamma$-centered pocket.
However, since this pocket is small (with a cross-sectional area on the order of 1\% of the Brillouin zone) and has a relatively complex shape, the numerical agreement of the frequencies is not satisfactory.

It is noteworthy that the present calculation can explain branches $\beta$ and $\zeta$, which could not be explained by previous band-structure calculations \cite{Lee94PRB, Singh96SSC, Yamauchi04PhysicaC}.
The band-26 sheet surrounding the $\Gamma$Z line looks like a flower with four petals.
In the present calculation, the four petals are connected on the Z plane ($k_z$ = 2$\pi/c$), which allows orbits $\beta$ and $\zeta$ [Fig. 3, see also Figs. S3(b) and (c)].
In the previous calculation of \cite{Yamauchi04PhysicaC} (top right inset of Fig. 3), the four petals are not connected on the Z plane, and there are gaps between adjacent petals.

\begin{figure}
\includegraphics[width=7cm]{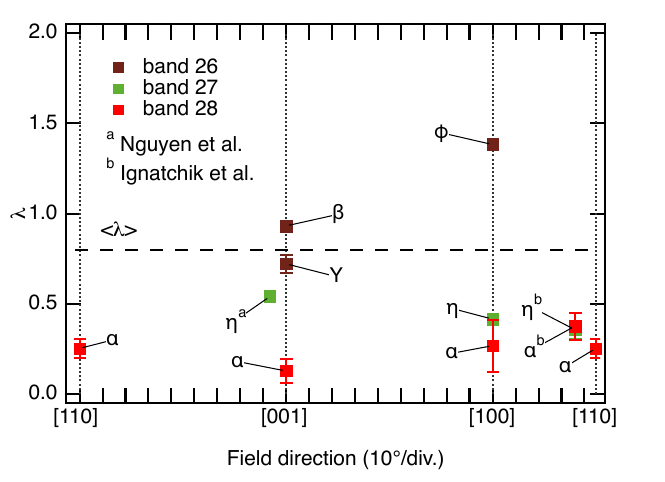}
\caption{\label{Fig4}Mass enhancement parameter $\lambda = m^*/m_{band} - 1$ in YNi$_2$B$_2$C.
The marks labelled with a and b are based on the effective masses reported in \cite{Nguyen96JLTP, Ignatchik05JMMM}.
The horizontal dashed line indicates the Fermi-surface averaged mass enhancement $\langle \lambda \rangle$.
}
\end{figure}

We determined effective masses for frequencies observed for $B \parallel$ [110], [001], and [100] by measuring temperature dependence of the oscillation amplitudes \cite{SM}.
We then determined the mass enhancement parameters $\lambda = m^*/m_{band} - 1$ by comparing the experimental effective masses and theoretical band masses (see Table SI).
We plotted the obtained mass enhancement parameters in Fig. 4, where we also plotted values estimated from the effective masses reported in \cite{Nguyen96JLTP, Ignatchik05JMMM} together (labelled with a and b).
$\lambda$ varies by about an order of magnitude, from 0.13(7) for the $\alpha$ orbit with $B \parallel c$ to 1.38(2) for the $\phi$ orbit with $B \parallel a$. 
This indicates a significantly larger variation in comparison with MgB$_2$, where $\lambda$ differs at most by a factor of three: $\sim$0.4 for the $\pi$ band and $\sim$1.1 for the $\sigma$ band \cite{Carrington03PRL}.
%The large variation in the electron-phonon coupling is generally consistent with the multigap superconductivity in YNi$_2$B$_2$C suggested by previous studies \cite{Shulga98PRL, Nohara99JPSJ, Boaknin01PRL}.

Figure 4 clearly shows the Fermi-surface-sheet dependence of the mass enhancement.
The mass enhancement decreases in the order of the band-26 sheet, band-27 P-centered sheet, and band-28 sheet.
The improved band-structure calculation in the present study enabled us to observe this trend clearly.
The $\lambda$ values for the band-27 P-sheet and the band-28 sheet are significantly smaller than the Fermi-surface average $\langle \lambda \rangle$ (horizontal dashed line) determined from specific heat measurements, suggesting that these parts of the Fermi surface may account for the multiband superconductivity observed in this compound through various experiments \cite{Shulga98PRL, Nohara99JPSJ, Boaknin01PRL}.
Band-27 constitutes approximately 20\% of the total density of states (DOS) at the Fermi level, although the contribution of band-28 is minor. 
However, it is difficult to confirm the pointlike minimum of the gap reported by ARPES \cite{Baba10PRB}.

Previous results on LuNi$_2$B$_2$C did not show such a clear trend \cite{Bergk08PRL}:
The mass enhancements of the cushion and sphere sheets in \cite{Bergk08PRL}, which correspond to our band-27 P-sheet and band-28 sheet, respectively, showed a crossover depending on the magnetic field direction.
The mass enhancement for the branched FS in \cite{Bergk08PRL}, which corresponds to our band-26 sheet, was no larger than that for the cushion sheet (corresponding to our band-27 P-sheet).
In addition, the mass enhancement for the cube sheet in \cite{Bergk08PRL}, which corresponds to our band-27 $\Gamma$-centered sheet, was reported to largely vary with the field direction: it was $\sim$1 for [100] but became $\sim$2.7 for [001] \cite{Bergk08PRL}.
The lack of a clear trend in the mass enhancement in \cite{Bergk08PRL} might reflect the inadequate agreement between their calculated Fermi surface and the experimental dHvA results.
Specifically, their Fermi surface fails to explain the $\beta$ and $\zeta$ frequencies, and its calculated frequencies for the $\alpha$ orbit show poor agreement with the experimental data \cite{Bergk08PRL}.

Figure 4 shows that the mass enhancement for the band-28 electron sheet is very weak.
Especially, it is only 0.1(1) for the field direction of [001].
This is very consistent with the conclusion drawn from the observation of dHvA oscillations from this sheet far below $B_{c2}$ for $B \parallel c$ that the gap on this   
sheet is much smaller than gaps on other sheets \cite{Terashima97PRB, Nossler17PRB}.
It is noteworthy that the two independent methodologies based on dHvA measurements, i.e. estimation of the mass enhancement (present study) and observation of dHvA oscillations in the mixed state \cite{Terashima97PRB, Nossler17PRB}, reached the same conclusion that the band-28 sheet is passive with respect to superconductivity in YNi$_2$B$_2$C.

The mass enhancement due to the electron-phonon coupling in YNi$_2$B$_2$C was theoretically studied in \cite{Kawamura17PRB}.
The largest electron-phonon coupling was found on the band-27 $\Gamma$-centered sheet.
Unfortunately, we were unable to identify dHvA frequencies arising from this sheet with certainty in the present study.
The study \cite{Kawamura17PRB} indicated that the second most strongly coupled sheet was the band-26 sheet, while the third the band-27 P-sheet.
This is consistent with the present result (Fig. 4).
For the band-28 sheet, the band-structure calculation in \cite{Kawamura17PRB} wrongly predicted two closed pockets around $\Gamma$, and hence comparison cannot be made with the present result.

In summary, we performed dHvA measurements on YNi$_2$B$_2$C.
Our improved band structure calculation successfully explained most of observed frequency branches.
Especially, it revealed the origins of the previously inexplicable large frequencies $\beta$ and $\zeta$.
We determined the many-body mass enhancement parameter $\lambda$ by comparing experimental effective masses with band masses.
The results showed a clear Fermi-sheet dependence of the electron-phonon coupling that the coupling weakens in the order of the band-26 sheet, band-27 P-sheet, and band-28 sheet.
The electron-phonon coupling for the band-28 sheet is very weak, almost absent for $B \parallel c$.
This strongly suggests that the superconducting gap on this sheet is small and hence is in line with the previous reports that dHvA oscillations from this sheet can be observed deep in the mixed \cite{Terashima97PRB, Nossler17PRB}.
There was a certain level of agreement between the electron-phonon coupling experimentally deduced in this study and that theoretically calculated in \cite{Kawamura17PRB}.
Considering recent discoveries of electron-phonon coupling high-transition-temperature superconductivity in hydrides \cite{Drozdov15Nature, Drozdov19Nature}, accurate theoretical calculations of electron-phonon coupling are of significance.
This study provides essential input for the development of such methods. 
In addition, our results clearly demonstrate that strong electron-phonon coupling can be realized in compounds containing light elements such as carbon or boron, even when heavier elements like Ni and Y are present. 
This provides an important guideline for the future exploration of superconducting materials.

\begin{acknowledgments}
TT thanks Haruyoshi Aoki, Shinya Uji, Chieko Terakura, and Kazuo Kadowaki for their supports and encouragements during this study.
TT also thanks Tamio Oguchi and Kunihiko Yamauchi for valuable discussions.
This work was supported by JSPS KAKENHI Grant Numbers JP22K03537 and JP24K00587.
MANA is supported by World Premier International Research Center Initiative (WPI), MEXT, Japan.
\end{acknowledgments}

\end{document}

\preprint{ver4SM}

\title{Fermi-surface-sheet dependent electron-phonon coupling in a borocarbide superconductor YNi$_2$B$_2$C}

\author{Taichi Terashima}
\email{TERASHIMA.Taichi@nims.go.jp}
\affiliation{Research Center for Materials Nanoarchitectonics (MANA), National Institute for Materials Science, Tsukuba 305-0003, Japan}
\author{Hiroyuki Takeya}
\affiliation{Research Center for Energy and Environmental Materials (GREEN), National Institute for Materials Science, Tsukuba 305-0047, Japan}
\author{Hisatomo Harima}
\affiliation{Department of Physics, Kobe University, Kobe 657-8501, Japan}

\date{\today}

\maketitle

\renewcommand{\figurename}{FIG. S}
\renewcommand{\tablename}{TABLE S}

\renewcommand{\bibnumfmt}[1]{[SM#1]} 
\renewcommand{\citenumfont}[1]{SM#1}

\section{Crystal growth and characterization}
Single crystals of YNi$_2$B$_2$C were grown by a floating zone method without a crucible to avoid impurity inclusions into grown crystals, as detailed in \cite{Takeya96PhysC}. 
The crystals were examined by SEM, EPMA and X-ray diffraction technique and found to be of high quality without any secondary phase.
Details of the results of these characterizations can be found in Refs. \cite{Takeya96PhysC, Takeya96JAC, Takeya96JMMM} (the dHvA measurements reported in the present paper were performed between 1995 and 1997, in parallel with these characterizations).
The lattice parameters were $a$ = 3.5265(1) and $c$ = 10.5415(3) \AA, which are in excellent agreement with those reported in \cite{Godart95PRB}.
The superconducting transition was observed at $T_c$ = 15.6 K (onset) with a very narrow transition width of 0.7 K \cite{Takeya96PhysC}.
The residual resistivity ratio RRR was 29.3 with electrical current applied along the $a$-axis, and the resistivity at $T_c$ was 2.89 $\mu\Omega$cm. 

\section{de Haas--van Alphen measurements}
The reported dHvA measurements were performed with the field-modulation technique \cite{Shoenberg84, Terashima13PRBdHvA}, and detection was made at the second harmonic of the modulation frequency.
The modulation frequency and amplitude were typically 67 Hz and 10 mT, respectively.
The insets of Fig. 2 (main text) show examples of raw data without background subtraction.
The effective mass $m^*$ for each frequency (orbit) was determined by fitting the temperature reduction factor $R_T$ of the Lifshitz-Kosevich formula to the experimental temperature dependence of the oscillation amplitude, which was determined by Fourier transformation:
$R_T= X/\sinh X$, where $X=rK\mu^* T/B$ for a fundamental ($r=1$) or a $r$-th harmonic frequency, $\mu^*=m^*/m_e$, and the coefficient $K$ is 14.69 T/K \cite{Shoenberg84}.
Figures S1 and S2 show examples of mass determination.

\begin{figure}
\includegraphics[width=17cm]{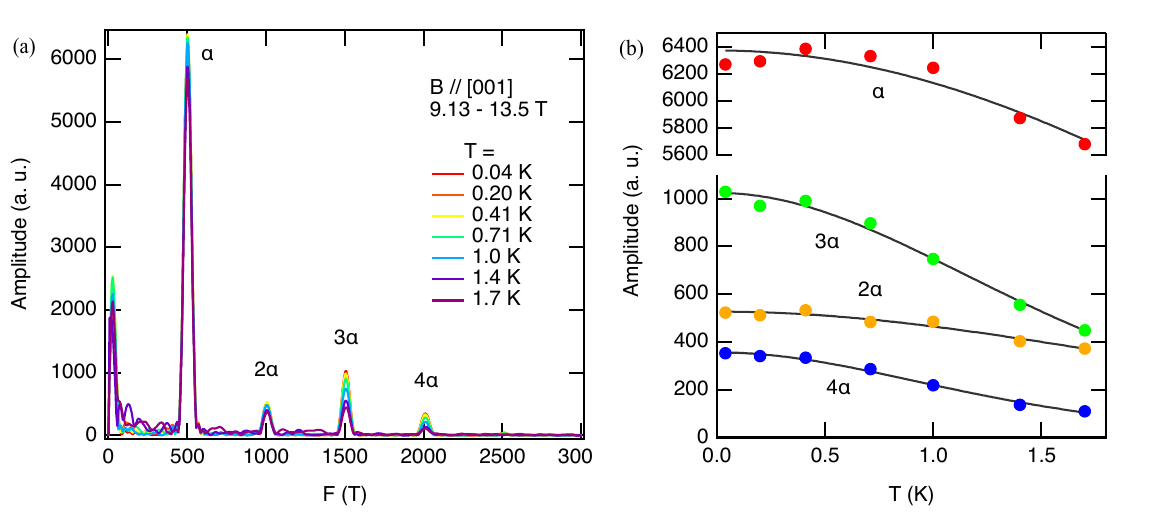}
\caption{\label{FigS1}Effective mass of $\alpha$.
(a) Fourier transforms of oscillations measured with $B \parallel c$ at indicated temperatures.
(b) Temperature dependences of oscillation amplitudes for $\alpha$, $2\alpha$, $3\alpha$, and $4\alpha$.
The solid lines are fits to $R_T$, which yield $m^*/m_e$ = 0.36(3), 0.33(2), 0.35(1), and 0.33(1), respectively, where the errors in parentheses are numerical fitting errors.
From the weighted average, we conclude $m^*/m_e$ = 0.34(2).
}
\end{figure}

\begin{figure}
\includegraphics[width=17cm]{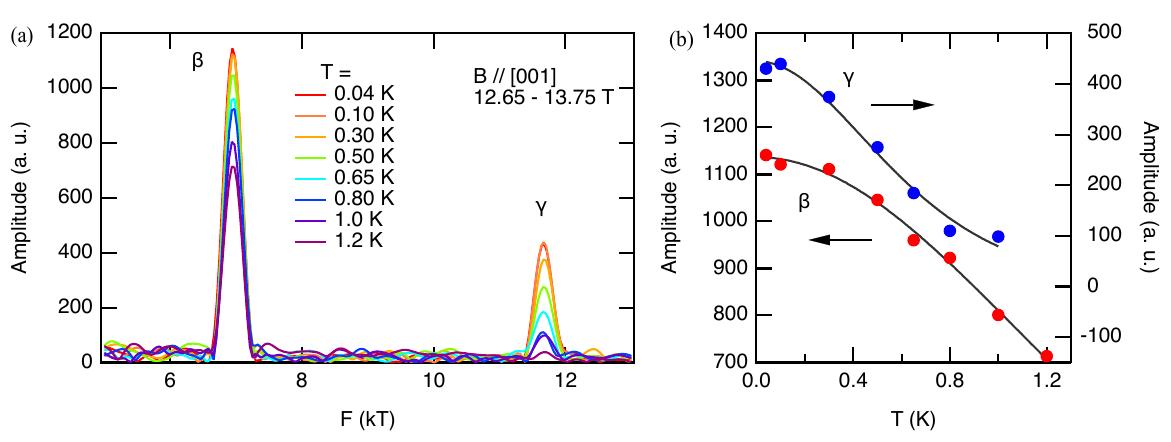}
\caption{\label{FigS2}Effective masses of $\beta$ and $\gamma$.
(a) Fourier transforms of oscillations measured with $B \parallel c$ at indicated temperatures.
(b) Temperature dependences of oscillation amplitudes for $\beta$ and $\gamma$.
The solid lines are fits to $R_T$, which yield $m^*/m_e$ = 1.32(3) and 3.3(1), respectively, where the errors in parentheses are numerical fitting errors.
}
\end{figure}

\section{Band-structure calculations}
The band-structure calculations are carried out by using FLAPW method based on the local density approximation (LDA) formula proposed by Gunnarsson and Lundqvist \cite{Gunnarsson76PRB}.
As mentioned in the main text, we shifted the Y $d$ and Ni $d$ levels upward from the LDA levels by 0.11 and 0.05 Ry, respectively.
The method of the calculation and the parameters used are the same as those in the previous one \cite{Yamauchi04PhysicaC}, except the number of sampling $k$-points.
The experimentally observed lattice constants are used: $a$ = 3.526 \AA, $c$ = 10.543 \AA, and $z$(B) = 0.358 \cite{Godart95PRB}.
Muffin-tin (MT) radii are set as 0.1747$a$, 0.1358$a$, 0.1010$a$, and 0.1010$a$ for Y, Ni, B, and C respectively.
Core electrons (Kr-core minus 4$p^6$ for Y, Ar-core minus 3$p^6$ for Ni, He core for B and C) are calculated inside the MT spheres.
4$p^6$ electrons on Y and 3$p^6$ electrons on Ni are calculated as valence electrons by using the second energy window.
The scalar relativistic effects are taken into account for all electrons, and the spin-orbit coupling is considered self-consistently for all valence electrons inside the MT spheres as in a second variational procedure.
The LAPW basis functions are truncated at $|\mathbf{k}+\mathbf{G}_i|= 4.01 \times 2\pi/a$ corresponding to 403 LAPW functions at the $\Gamma$ point.

The sampling 767 $k$-points (divided by 24, 24, and 8) are uniformly distributed in the irreducible 1/16th of the bct Brillouin zone (BZ), instead of 369 k-points (divided by 16, 16, and 8) in the previous work \cite{Yamauchi04PhysicaC}.
Accordingly, the energies are calculated at about twice as many points within each $k_z$ = constant plane compared to the previous study.
To draw the perspective view of the Fermi surfaces and calculate the extremal areas, the energy eigenvalues are interpolated at the 226,981 mesh points.

\begin{figure}
\includegraphics[width=17cm]{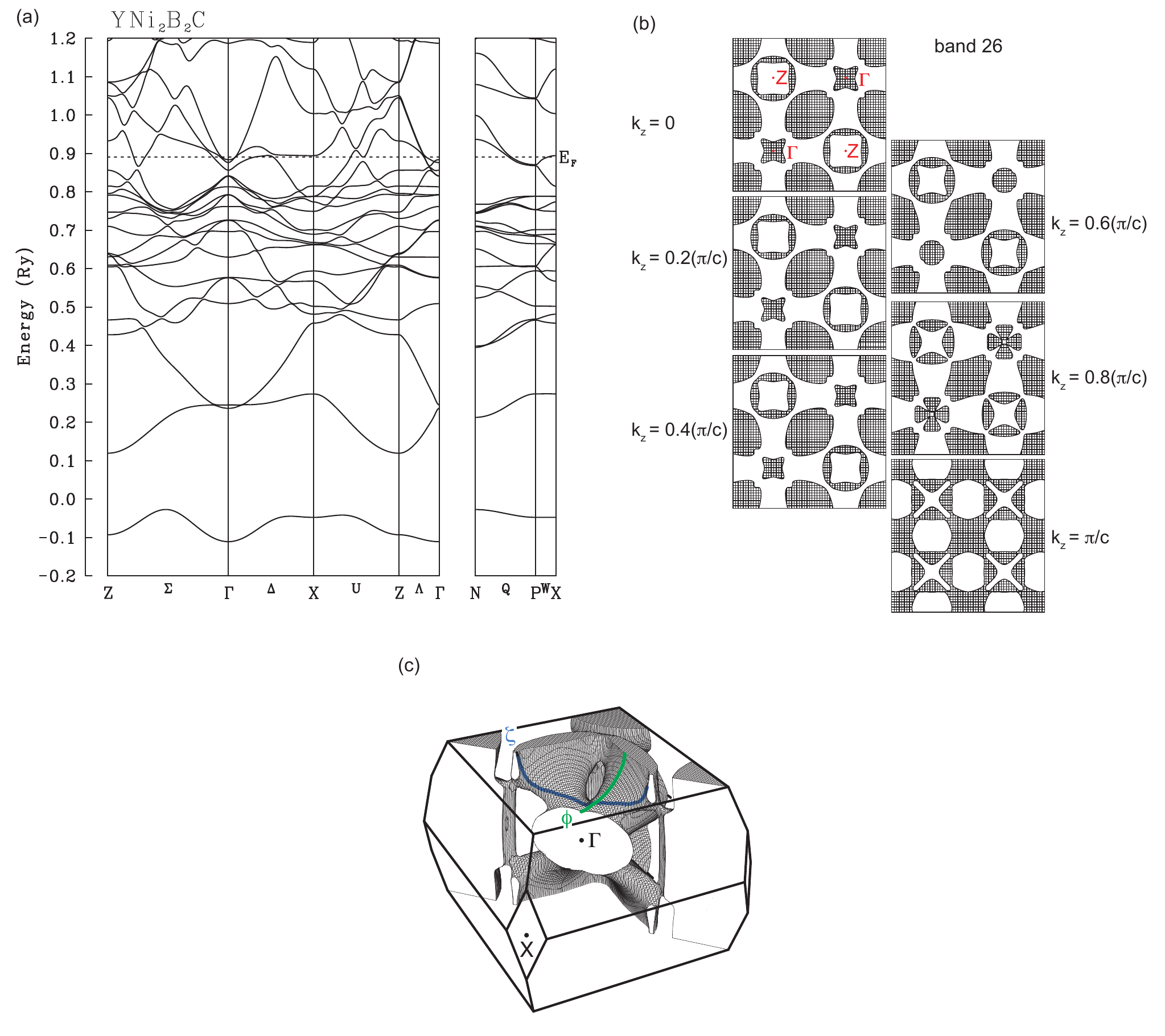}
\caption{\label{FigS1}(a) Calculated band structure.
(b) Cross-sections of the band-26 Fermi surface.
(c) $\zeta$ and $\phi$ orbits on the band-26 Fermi surface.
}
\end{figure}

\section{Estimation of $\lambda$}
Table S1 provides the source data for the mass enhancement factor $\lambda$ shown in Fig. 4.

\begin{table}%[H] add [H] placement to break table across pages
\caption{\label{Tab1}Experimental and calculated dHvA frequencies and masses, and mass enhancements calculated as $\lambda = m^*/m_{band}-1$.  The errors quoted for the effective masses are numerical fitting errors, from which the errors for $\lambda$ were derived.}
\begin{ruledtabular}
\begin{tabular}{ccccccc}
\multicolumn{2}{c}{} & \multicolumn{2}{c}{Experiment} & \multicolumn{2}{c}{Calculation} & \\
\cline{3-4} \cline{5-6}
\multicolumn{1}{c}{$\theta$ ($^{\circ}$)} & Branch & \multicolumn{1}{c}{$F$ (kT)} & \multicolumn{1}{c}{$m^*/m_e$} & \multicolumn{1}{c}{$F$ (kT)} & \multicolumn{1}{c}{$m_{band}/m_e$} & \multicolumn{1}{c}{$\lambda$} \\
\hline
{[110]} & $\alpha$ & 0.79 & 0.48(2) & 0.84 & 0.38 & 0.25(5) \\
\\
{[001]} & $\gamma$ & 11.7 & 3.3(1) & 11.9 & 1.9 & 0.72(5) \\
{[001]} & $\beta$ & 7.0 & 1.32(2) & 6.4 & 0.68 & 0.93(3)\\
{[001]} & $\alpha$ & 0.50 & 0.34(2) & 0.50 & 0.30 & 0.13(7)\\
\\
{[100]} & $\phi$ & 4.5 & 1.27(1) & 4.2 & 0.53 & 1.38(2)\\
{[100]} & $\eta$ & 1.8 & 1.86(2) & 1.7 & 1.32 & 0.41(2)\\
{[100]} & $\alpha$ & 0.83 & 0.53(6) & 0.87 & 0.42 & 0.3(1) \\
\\
$\theta_{(1\bar{1}0)}=7^{\circ}$\footnotemark[1]  & $\eta$ & 5.5 & 3.2 & 4.8 & 2.1 & 0.54\\
\\
$\varphi=36^{\circ}$\footnotemark[2]  & $\eta$ &  & 2.7(1) & 2.1 & 2.0 & 0.35(5)\\
$\varphi=36^{\circ}$\footnotemark[2]  & $\alpha$ &  & 0.54(3) & 0.84 & 0.39 & 0.37(8) \\

% Lines of table here ending with \\
\end{tabular}
\end{ruledtabular}
\footnotetext[1]{From \cite{Nguyen96JLTP}.  The field direction is 7$^{\circ}$ from [001] towards [110].  The frequency was read from Fig. 2 of \cite{Nguyen96JLTP}. }
\footnotetext[2]{From \cite{Ignatchik05JMMM}.  The field direction is 36$^{\circ}$ from [100] towards [110].  The frequency values were not given in \cite{Ignatchik05JMMM}.}
\end{table}

%\bibliography{/Users/tera/GoogleDrive/Work/bib_files/literature.bib}